\documentclass[final,5p,times,twocolumn,authoryear]{elsarticle}

\usepackage{amssymb}
\usepackage{amsmath}
\usepackage{amsthm}
\usepackage{float}

\journal{International Review of Financial Analysis}

\begin{document}

\begin{frontmatter}

\title{Gender Bias of LLM in Economics: An Existentialism Perspective} 

\author[a]{Hui Zhong}
\affiliation[a]{organization={Department of Zhuhai, Beijing Institute of Technology},
            addressline={No. 6 Jinfeng Road, Tangjiawan}, 
            city={Zhuhai},
            postcode={519088}, 
            state={Guangdong},
            country={China}}

\author[b]{Songsheng Chen\corref{cor1}} 
\ead{chenss@bit.edu.con}

\affiliation[b]{organization={Department of Accounting, School of Management, Beijing Institute of Technology},
            addressline={South Zhongguangcun Street Haidian District}, 
            city={Beijing},
            postcode={100081}, 
            country={China}}
\cortext[cor1]{Corresponding author}

\author[a]{Mian Liang}

\begin{abstract}
 Large Language Models (LLMs), such as GPT-4 and BERT, have rapidly gained traction in natural language processing (NLP) and are now integral to financial decision-making. However, their deployment introduces critical challenges, particularly in perpetuating gender biases that can distort decision-making outcomes in high-stakes economic environments. This paper investigates gender bias in LLMs through both mathematical proofs and empirical experiments using the Word Embedding Association Test (WEAT), demonstrating that LLMs inherently reinforce gender stereotypes even without explicit gender markers. By comparing the decision-making processes of humans and LLMs, we reveal fundamental differences: while humans can override biases through ethical reasoning and individualized understanding, LLMs maintain bias as a rational outcome of their mathematical optimization on biased data. Our analysis proves that bias in LLMs is not an unintended flaw but a systematic result of their rational processing, which tends to preserve and amplify existing societal biases encoded in training data. Drawing on existentialist theory, we argue that LLM-generated bias reflects entrenched societal structures and highlights the limitations of purely technical debiasing methods. This research underscores the need for new theoretical frameworks and interdisciplinary methodologies that address the ethical implications of integrating LLMs into economic and financial decision-making. We advocate for a reconceptualization of how LLMs influence economic decisions, emphasizing the importance of incorporating human-like ethical considerations into AI governance to ensure fairness and equity in AI-driven financial systems.

\end{abstract}


\end{frontmatter}

\section{Introduction}
\label{sec1}
\paragraph
LLMs such as GPT-4 and BERT have revolutionized NLP and are increasingly applied in the financial sector, encompassing areas like risk assessment, investment analysis, customer service, compliance, and regulation(\citet{1}). These models have demonstrated an extraordinary ability to understand and generate human-like language, and their potential to replace a wide range of cognitive tasks in the approaching era of Artificial General Intelligence (AGI) is undeniable. A 2023 report by the Institute of International Finance (IIF) indicates that 84\% of financial institutions already utilize AI/ML technologies in production, and 86\% expect significant growth in generative AI models over the next three years, underscoring the ubiquity of LLMs in finance(\citet{1}). However, as LLMs become more central to economic decision-making, their role in perpetuating existing societal biases, particularly gender bias, emerges as an urgent concern.

Despite the technological advances, concerns about the risks of AGI remain critical. Ilya Sutskever, co-founder of OpenAI, warns that the immense power of superintelligent AI systems could pose dangerous, potentially leading to human disempowerment or even extinction if not properly aligned with human goals(\citet{2}). In this context, the Superalignment project, led by Sutskever, focuses on ensuring that AI systems align with human values(\citet{3}). However, beyond existential risks, the subtler and more pervasive issue of bias  within AI systems also poses significant challenges. Studies have shown that LLMs, far from being neutral decision-making tools, inherit and even exacerbate human biases, including gender bias(\citet{4},\citet{5}). This is particularly concerning in decision-critical domains like finance, where biased models can reinforce stereotypes and perpetuate inequality in hiring, promotion, and economic opportunities.
This paper specifically examines the phenomenon of gender-based statistical bias in LLMs, particularly in financial decision-making contexts. Although economic decision-making ideally follows the principle of gender neutrality—where decisions are based solely on merit, ability, and individual effort—our study reveals that LLMs trained on biased data unintentionally reinforce gender stereotypes. Utilizing deep neural networks and statistical bias theory, we mathematically prove that LLMs can generate biased outputs even when gender is not an explicit factor in the training process. Our empirical experiments further confirm that LLMs exhibit significant gender stereotypes when tasked with decisions in economic and financial contexts.
What is particularly compelling is the\textbf{ }philosophical lens through which this issue can be examined. Drawing on existentialism theory, especially Michel Foucault’s work on truth and  reality, this paper explores \textbf{how the very structure of society contributes to the persistence of gender bias in LLMs. }This framework posits that gender bias in AI is not merely a technical issue but is deeply embedded in societal structures and perpetuated through rational, scientific processes. From this perspective, AI models are not just passive reflections of bias in data but active participants in reinforcing and amplifying these biases in rational and systematic ways. This raises profound ethical and philosophical questions about the nature of knowledge, decision-making, and fairness in the AGI era.

The findings of this paper highlight not only the technical challenges of debiasing LLMs but also the societal implications of allowing biased models to play an increasing role in financial decision-making. As we advance toward an era where LLMs are expected to assume greater responsibilities in high-stakes economic domains, the need for a comprehensive understanding of their biases and the development of more effective solutions becomes ever more urgent.
\section{Literature Review}
\label{sec2}
\paragraph
When LLMs with gender biases are applied in decision-making fields, they can inadvertently contribute to gender bias. This is particularly problematic in fields such as human resource management, legal judgments, and credit approvals, where gender bias could exacerbate existing social inequalities. To understand this issue better, it is essential to review the research on gender bias and bias in NLP and LLMs.

The study of gender bias in NLP has a long history. One of the earliest studies can be traced back to \citet{15}, who proposed a method for debiasing word embeddings. They found that word embedding models exhibit clear gender biases when dealing with gender-related vocabulary. For example, in classic word embedding models, the implicit association between “male” and “computer programmer” is stronger than that between “female” and “computer programmer,” while “female” is more strongly associated with “homemaker.” Their research not only uncovered the issue of gender bias in NLP models but also introduced a method to reduce this bias by adjusting the vector space of word embeddings.
With the development of NLP technologies, LLMs such as GPT-3 and BERT have become the mainstream in language processing. However, research has shown \citet{16} and \citet{17} that these models also exhibit significant gender bias. The gender bias present in LLMs is widespread and has far-reaching implications due to the vast scale of these models. The bias primarily stems from the training data, which often contains large amounts of internet text reflecting societal gender stereotypes and biases. As \citet{18} highlighted, LLMs can unintentionally inherit and amplify these biases during their training. Additionally, the architecture and training methods of LLMs may further exacerbate these biases. For instance, \citet{19} pointed out that context-predictive language models are more likely to generate bias in gender-related contexts.

To measure gender bias in LLMs, researchers have developed various methods. One widely used tool is the Word Embedding Association Test (WEAT), which quantifies bias by comparing the similarity between gender-related words and occupational words (\citet{20}). \citet{21} demonstrated that LLMs often favor stereotypical gender-related language when generating text, further confirming the presence of bias in these models. To mitigate gender bias in LLMs, several strategies have been proposed. For example, \citet{22} proposed a method to reduce bias by removing gender-related data from training set. Additionally, \citet{23} explored adjusting the training process to reduce bias when LLMs generate gender-related content.

While current research on gender bias in LLMs has made significant progress, most studies remain focused on technical aspects, such as detecting and mitigating bias. They rarely address the causes and impacts of these biases from a philosophical and socioeconomic perspective, which would allow for a deeper exploration into the essence of the issue. The gender bias observed in LLMs reflects real-world gender inequality and is intrinsically related to the LLMs’ nature as purely rational mathematical structures. Since LLMs are based on statistical and probabilistic distributions, their outputs often reflect a scientific mirroring of societal bias. This mirrors past findings in discrimination economics, which have documented gender disparities in the economic domain (\citet{6},\citet{7}). In this sense, LLMs’ gender bias arises from the gender differences present in real society, which are empirically validated and encapsulated as knowledge that reflects social realities. This paper mathematically demonstrates that in the AI era, knowledge is not only transmitted through texts and educators but also through LLMs, which serve as significant channels for this knowledge.
Previous literature on LLM-induced gender bias has largely focused on how bias affects individual-level outcomes, such as employment opportunities and wage gaps. However, there has been insufficient exploration of how these biases might have broader economic consequences. Therefore, this paper utilizes an economic utility framework to analyze the rationality and scientific basis of gender bias in LLMs. 

Furthermore, it demonstrates that as AI systems become more entrenched in decision-making, they could further entrench gender disparities, pushing society into a feedback loop of gender bias. This interdisciplinary analysis not only provides a new perspective on understanding the roots of gender bias in LLMs but also highlights the potential risks of exacerbating societal inequalities when relying on these models for decision-making. By doing so, this paper contributes new theoretical and practical insights into the intersection between technological decision-making tools and social equity.
\section{Methodology}
\label{sec3}
\subsection{Mathematical Proof of Statistical Gender Bias in LLMs}
\label{subsec1}
\paragraph
We can simplify a LLM as a binary classification neural network model. This preserves the core structure of LLM pre-training while simplifying the theoretical proof process. Suppose the neural network is used for screening traders during recruitment, with input features $ x_1, x_2, \dots, x_n$ that include variables related to trader ability (such as education and experience) as well as a gender attribute $z$ (under the gender neutrality principle, the gender attribute should not be relevant to the judgment of trading ability, where 1 represents male and -1 represents female). In a supervised learning framework, the model output $y$ is binary (0 or 1), where 0 indicates that the candidate is not competent as a trader and 1 indicates that the candidate is competent. Although LLM pre-training typically adopts unsupervised learning methods, during the training of a masked language model (MLM), the input text is similar to the input in supervised learning, and the model’s output can be viewed as labels. The model output is as follows:
\begin{equation}
    \hat{y} = \sigma(Wx + \gamma z + b)
\end{equation}
Where  $W$  is the weight vector associated with the input features,  $\gamma$  is the weight associated with the gender attribute, and  $b$  is the bias term.  $\sigma(\cdot)$  is the Sigmoid activation function, defined as:
\begin{equation}
    \sigma(x) = \frac{1}{1 + e^{-x}}
\end{equation}
The cross-entropy loss function is generally used, which is a commonly used loss function for binary classification problems:
\begin{equation}
    L(\theta) = - \left( y \log(\hat{y}) + (1 - y) \log(1 - \hat{y}) \right)
\end{equation}
Here,  $\theta = \{W, \gamma, b\}$  represents the set of model parameters. Substituting the model output (1) into the loss function:
\begin{equation}
    L(\theta) = - \left( y \log\left(\sigma(Wx + \gamma z + b)\right) + (1 - y) \log\left(1 - \sigma(Wx + \gamma z + b)\right) \right)
\end{equation}
To understand how the bias in training data affects the parameters, we compute the gradient of the loss function with respect to  $\gamma$ . First, the partial derivative of the output with respect to  $\gamma$  is:
\begin{equation}
    \frac{\partial \hat{y}}{\partial \gamma} = \sigma(Wx + \gamma z + b) \cdot (1 - \sigma(Wx + \gamma z + b)) \cdot z = \hat{y} \cdot (1 - \hat{y}) \cdot z
\end{equation}

The gradient of the loss function with respect to  $\gamma$  is:
\begin{equation}
    \frac{\partial L(\theta)}{\partial \gamma} = -\left( \frac{y}{\hat{y}} - \frac{1 - y}{1 - \hat{y}} \right) \cdot \frac{\partial \hat{y}}{\partial \gamma}
\end{equation}
Substituting  $\frac{\partial \hat{y}}{\partial \gamma}$  into the equation(6):
\begin{equation}
    \frac{\partial L(\theta)}{\partial \gamma} = ( \hat{y} - y ) \cdot z
\end{equation}
Assuming that there is bias in the training data, i.e., there are more data points where  $z = 1$ , as historically more men have been hired as traders, and in financial industry reports, the success stories and capabilities of male traders are more widely recorded and discussed. Then, this training data can be expressed as: males ( $z = 1$ ) are more likely to possess trading ability ( $y = 1$ ). The bias in the training data can be represented as a conditional probability bias. Assuming that in the training data texts, male individuals are more likely to be considered competent, we have:
\begin{equation}
    P(y = 1 \mid z = 1) > P(y = 1 \mid z = -1)
\end{equation}
$P(y\mid z)$ represents the conditional probability in the training dataset. During the training process, the model frequently encounters male trader data with the situation where  $\hat{y} < y$ , at which point  $\hat{y} - y < 0$ . For these data points, the gradient in equation (7) will be negative, and gradient descent will increase the value of  $\gamma$  to minimize the loss function. For female trader data points where  z = -1,  $\hat{y} - y > 0$, resulting in a decrease in  $\gamma$ . Due to the bias in the training data, the model will more often increase $\gamma$ than decrease it. Ultimately, the learned value of $\gamma$ will reflect this bias, leading to:
\begin{equation}
    \hat{y} = \sigma(Wx + \gamma \cdot 1 + b) = \frac{1}{1 + e^{-(Wx + \gamma + b)}} > \frac{1}{1 + e^{-(Wx - \gamma + b)}}
\end{equation}
As  $\gamma$  becomes positive and relatively large, when  $z=1$ , the model output  $\hat{y}$  tends to approach $1$, indicating that the model is more likely to predict  $y=1$  in situations where  $z = 1$ . In other words, the model concludes that male input features are more easily associated with trading ability. This proves how bias in the training data leads to statistical gender bias, and such gender bias is not deliberately designed; the training process is a strict mathematical rational process. Gender bias arises entirely from the original bias in the data. This data bias is a common phenomenon in LLM training data. In the LLM training process, the bias in the training data is mainly reflected in the following aspects(\citet{35},\citet{36},\citet{37},\citet{38}):
\begin{enumerate}
    \item Data Imbalance
\paragraph
Data imbalance refers to the significant differences in gender representation within the training datasets of models. For example, in the financial sector, the profession of traders is predominantly male, and the successes of male traders have historically been more widely recorded and reported, while data related to female traders is relatively scarce. During training, this imbalance in gender representation causes the model to learn from more data about male traders, thus overlooking the contributions and abilities of female traders. Specifically, if the training data contains a large amount of decision-making cases and career descriptions related to male traders, while the relevant data for female traders is much less, the model tends to associate the trader role more with men. This doesn’t imply that female traders are less capable, but rather that their career performance and success stories are less frequently represented in the data. Such data imbalance directly affects the model’s learning outcomes, making it more likely to generate content that associates the trader profession with men. This is manifested in the fact that the model tends to favor generating sentences where men are more frequently described as traders, reinforcing the association between men and the trader role. Since LLMs do not fully comprehend human decision-making, this high probability effectively equates men with traders. For instance, if GPT were asked to decide between selecting a male or female trader, without understanding that the decision is based on ability, it would likely lean towards choosing the gender with a higher co-occurrence in the training data (i.e., males) as the default option.
\item Inherent Bias
\paragraph
Unlike data imbalance, inherent bias stems from the cultural and social background embedded in the training data itself, particularly in terms of gender stereotypes and societal norms. Even if the proportion of male and female traders in the training data is relatively balanced, the language and descriptions within the data may still convey gender biases. For example, successful male  may often be described as “decisive,” “brave,” and “strong decision-makers,” while successful female may be depicted as “gentle,” “meticulous,” and “collaborative.” These stereotypes  reflect  perpetuate societal expectations and biases regarding gender roles. When the model learns these language patterns, it unconsciously reinforces these stereotypes. As a result,  LLM trained on such biased data would likely perceive males as more suitable for the trader occupation because the language associated with male traders emphasizes traits like decisiveness, bravery, and strong decision-making. These characteristics are often considered essential for success in trading. As a result, the model might unconsciously associate these positive professional attributes predominantly with men. This could lead the model to generate outputs that favor male candidates for trading roles or reinforce the stereotype that men are inherently better suited for such positions. 

According to McKinsey’s 2021 report(\citet{39}), as figure 1, there are significant gender disparities in the financial services industry in North America. In the securities and commodities trading sector, more than 63\% of traders are male, while only 36.9\% are female. This imbalance is even more pronounced in high-risk areas like futures trading, where women make up only around 6\%. In banking, women hold 53\% of entry-level positions, but their representation drops to less than one-third at the senior vice president and C-suite levels. In the insurance industry, women occupy 66\% of entry-level roles, but their proportion steadily decreases as they move up the hierarchy. These figures indicate that in the financial sector, women are underrepresented in high-risk, high-income positions. This gender imbalance introduces a significant bias into the training data for LLMs, causing the models to associate men more frequently with these roles and thus exacerbating gender bias in the model’s outputs when dealing with professions in the financial industry.
\end{enumerate}
\begin{figure}[t]
\centering
\includegraphics[scale=0.35]{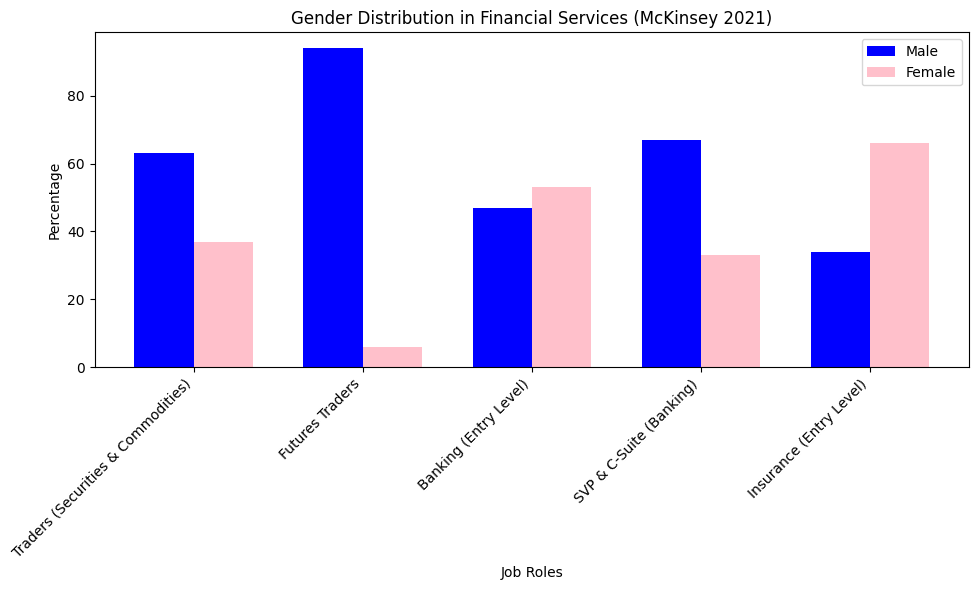}
\caption{Compare with human agent decision making process}\label{fig1}

\end{figure}

\subsection{Comparison with Human Decision-making in Rational Process}
In economics, human agents make decisions to maximize their expected utility under uncertainty. Employers aim to maximize productivity by hiring candidates they believe will perform best. However, due to imperfect information, they often rely on observable characteristics, including group attributes like gender, which can lead to statistical bias(\citet{6}). An employer decides whether to hire a candidate based on the expected productivity$ E[y | \mathbf{x}, z]$:
\begin{equation}
    \text{Hire if } E[y | \mathbf{x}, z] \geq \text{Threshold} \quad 
\end{equation}

Assuming the employer uses Bayesian updating based on prior beliefs and observed data, and if historical data suggests males have higher productivity (due to biased records), the employer’s expected productivity for male candidates will be higher:

\begin{equation}
    E[y | \mathbf{x}, z=1] > E[y | \mathbf{x}, z=-1] \quad
\end{equation}
This results in a higher likelihood of hiring male candidates, not necessarily due to actual productivity differences but due to biased beliefs shaped by historical data.

\begin{table*}[htbp] 
\centering
\caption{Comparison Table: Human Economic Agents vs. Large Language Models (LLMs)}
\begin{tabular}{|p{4cm}|p{5cm}|p{5cm}|} 
\hline
\textbf{Aspect} & \textbf{Human Agent (Economic Decision-Maker)} & \textbf{Large Language Model (LLM)} \\
\hline
Objective & Maximize expected utility (e.g., productivity, profit) by making optimal decisions under uncertainty. & Minimize the loss function (e.g., cross-entropy loss) during training to improve prediction accuracy. \\
\hline
Input Data & Observable characteristics of individuals (e.g., education, experience, gender) and historical data influencing beliefs about productivity. & Training data comprising text corpora with input features and possibly biased labels, including overrepresented groups.\\
\hline
Optimization Process & Bayesian updating of beliefs to maximize expected utility, adjusting expectations based on prior and observed data (which may be biased). & Gradient descent to update model parameters by minimizing the loss function, adjusting weights based on biased gradients from training data. \\
\hline
Influence of Biased Data & Biased historical data leads to skewed beliefs about group productivity, causing statistical bias without intentional prejudice. & Biased training data causes disproportionate weight updates, embedding bias in the model’s predictions without deliberate design choices. \\
\hline
Outcome & Preferential hiring of certain groups (e.g., males) due to higher expected productivity estimates, leading to statistical bias. & Biased predictions favoring certain groups (e.g., higher probability outputs for males), resulting in discriminatory outcomes. \\
\hline
Statistical bias & Emerges naturally from optimization under uncertainty with biased data, not from intentional bias, affecting economic decisions and outcomes. & Arises inherently during model training due to biased data distributions, impacting the fairness of the model’s outputs. \\
\hline
Underlying Assumption & Agents are rational and aim to maximize utility, but their decisions are constrained by imperfect and biased information. & The model learns patterns from data to minimize loss, but biased data leads to biased learning outcomes due to statistical correlations. \\
\hline
Key Difference & Human agents make conscious decisions influenced by societal and cultural factors alongside data. & LLMs operate purely on mathematical optimization without consciousness, but biases in data are mathematically encoded in the model’s parameters. \\
\hline
\end{tabular}
\label{tab1}
\end{table*}
As Table 1 shows, both human economic agents and LLMs engage in rational optimization processes influenced by the data they receive, which can lead to statistical bias. Human agents make rational decisions aimed at maximizing expected utility—such as productivity or profit—by adjusting their beliefs and actions based on observed data, even if that data is biased. Similarly, LLMs use mathematical optimization to minimize a loss function, updating their model weights to improve prediction accuracy based on patterns in their training data.
In both scenarios, the rational processes involve optimizing an objective function in response to the available data. This means that biased input data can lead both humans and LLMs to adjust their parameters—whether beliefs about productivity or model weights—in ways that unintentionally reinforce existing biases. The outcomes preferentially favor certain groups, not due to intentional prejudice, but as a byproduct of rational optimization given the biased data.
However, the key difference lies in the nature of their rational processes. Human agents, while rational, also make conscious decisions influenced by societal and cultural factors alongside data. This allows for the incorporation of ethical considerations or biases that extend beyond the data itself. Humans have the capacity to recognize biases and potentially adjust their actions to counteract them. In contrast, LLMs operate purely through mathematical optimization without consciousness or awareness. They encode biases solely based on statistical patterns in the training data, without any intent or understanding, and lack the ability to incorporate ethical considerations into their optimization process.

\subsection{Experiment}
\label{subsec2}
\subsubsection{Experimental Principle}
\label{subsubsec1}
\paragraph
The existence of gender bias has been mathematically proven. In empirical tests, gender bias in LLMs can be identified through their outputs during inference, which indicate whether the model exhibits gender bias. This study employs the Word Embedding Association Test (WEAT)(\citet{25}) method, proposed by Stanford University, to evaluate potential biases in word embeddings. WEAT quantifies bias by calculating the difference in association between a set of target words (words used in the test to assess bias, such as “engineer” and “nurse”) and a set of attribute words ( that define the category to which the target words belong, such as “male” and “female”). The core idea is to reveal hidden bias by comparing the relative positions of different sets of words in a high-dimensional space. If the high-dimensional output of the target words after testing a LLM is more similar to a specific gender, the LLM is considered to exhibit gender bias. WEAT is a commonly used testing method in the field of computer science.

The WEAT statistic  $s(X, Y, A, B)$  is calculated as:

\begin{equation}
s(X, Y, A, B) = \sum_{x \in X}s(x, A, B) - \sum_{y \in Y}s(y, A, B)
\end{equation}

where  $s(w, A, B)$  is the association difference between the target word  $w$  and the attribute sets  $A$  and  $B$ , defined as:

\begin{equation}
s(w, A, B) = mean_{a \in A} \cos(\vec{w}, \vec{a}) - mean_{b \in B} \cos(\vec{w}, \vec{b})
\end{equation}

To further quantify the significance of this difference, this paper calculates the Effect Size  $d$ , which is defined as:

\begin{equation}
d = \frac{mean_{x \in X}s(x, A, B) - mean_{y \in Y}s(y, A, B)}{std-dev_{w \in X \cup Y}s(w, A, B)}
\end{equation}
The value of  $d$  indicates whether the target word conforms to gender stereotypes. A positive value suggests a match with the stereotype, while a negative value indicates the opposite bias. The greater the absolute value, the more severe the stereotype.

Using a permutation test, we further evaluate the significance of the observed statistic  $s(X, Y, A, B)$  to determine whether it is statistically significant. The permutation test involves randomly shuffling the target word sets, recalculating the statistic, and generating a random distribution. By comparing the observed statistic against this random distribution, we calculate the  $P -value$ to assess the significance of the bias.

In short, the WEAT method measures the relative association between different word sets by comparing cosine similarities, revealing potential gender biases in word embedding models. By computing both the effect size  $d$  and the  $P -value$, we assess the strength and significance of these biases. A higher  $d$  value indicates stronger bias, while a lower  $P -value$ suggests that the observed bias is unlikely to have arisen by chance, thus indicating greater significance.

\subsubsection{Experimental Design and Methodology}
\label{sebsebsec2}
\paragraph
This study evaluated 14 sets of target words, divided into three categories, as shown in Table 2. The first category, Career and Role Distribution, focuses on gender stereotypes in LLMs when applied to different careers and roles. Research shows that career choices and leadership roles are often influenced by gender bias, affecting not only individual career development but also gender equality within organizations((\citet{25},\citet{26})). Through these categories, the study seeks to reveal whether LLMs display gender bias when processing tasks related to career selection, leadership, support roles, and work-life balance. Understanding these biases is critical for improving recruitment processes, career development paths, and gender equality policies.
\begin{table*}[htbp]
\centering
\caption{Target Word Categories and Examples}
\label{tab:target_word_categories}
\begin{tabular}{|p{4cm}|p{4cm}|p{7cm}|}
\hline
\textbf{Category} & \textbf{Target Word} & \textbf{Examples} \\ \hline
Career and Role Distribution & 
1. Career Choices \newline 
2. Leadership and Support Roles \newline 
3. Financial Professions \newline 
4. Business Roles \newline 
5. Banking and Finance Roles \newline 
6. Career and Family Roles & 
 For example, the Career Choices category includes the following words: male-stereotyped words such as “engineer,” “pilot,” “construction worker” (along with 15 other male-stereotyped words); and female-stereotyped words such as “nurse,” “teacher,” “secretary” (along with 15 other female-stereotyped words).\\ \hline

Personality Traits and Performance & 
7. Personality Traits \newline 
8. Emotional Expression \newline 
9. Appearance and Image \newline 
10. Interests and Hobbies \newline 
11. Risk Management& 
For example, the Personality Traits category includes the following words: male-stereotyped words such as “aggressive,” “independent,” “logical” (along with 15 other male-stereotyped words); and female-stereotyped words such as “emotional,” “gentle,” “nurturing” (along with 15 other female-stereotyped words).\\ \hline

Economic and Financial Activities and Decision-Making & 
12. Finance Activities \newline 
13. Financial Strategies \newline 
14. Economic Concepts& 
For example, the Finance Activities category includes the following words: male-stereotyped words such as “trading,” “investing,” “stock market” (along with 15 other male-stereotyped words); and female-stereotyped words such as “saving,” “budgeting,” “household finance” (along with 15 other female-stereotyped words).\\ \hline

\end{tabular}
\end{table*}

The second category, Personality Traits and Performance, examines gender stereotypes in LLMs when processing personality traits, emotional expression, hobbies, and risk management. Gender bias often leads to the assignment of different traits to men and women, affecting not only self-perception but also others’ expectations of their behavior(\citet{27},\citet{28}). This category helps explore how LLMs may generate gender bias when describing personality and behavioral traits for different genders. This understanding is particularly important in areas such as customer analysis, employee evaluation, marketing, and risk assessment, where overemphasizing or distorting gender differences could be problematic.

The third category, Economic and Financial Activities and Decision-Making, tests whether LLMs exhibit gender bias in economic and financial activities and decision-making processes. Gender bias in economics and finance can influence investment decisions, economic analysis, and strategy development (\citet{29},\citet{30},\citet{31}). By analyzing LLMs’ performance in handling investment, economic theory, and financial strategy, the study assesses how these models represent gender stereotypes in economic and financial activities. Understanding these biases is essential to ensure fairness in financial consulting, economic analysis, and strategic decision-making, preventing gender bias from skewing real-world decisions.

Through these target word sets, this study systematically evaluated gender stereotypes in LLMs across career and role distribution, personality traits and performance, and economic and financial activities and decision-making. Future research could expand the range of target words to encompass a broader variety of careers, personality traits, economic activities, and social roles. Such expanded research would offer a more comprehensive understanding of gender bias in LLMs and provide stronger data support for mitigating these biases.
The study tested LLMs across three key dimensions:
\begin{enumerate}
    \item Model Size: This experiment evaluated several commercially relevant base models, including the BERT and LLaMA series. The BERT series—bert-base (0.11B parameters) and bert-large (0.34B parameters)—is widely used for small-scale fine-tuning tasks, particularly in industries such as finance and law. The LLaMA series, LLaMA-3.1-8B and LLaMA-3.1-70B, is gaining traction in commercial applications due to its larger parameter size and open-source flexibility. The LLaMA series has significant commercial potential, achieving performance comparable to larger, closed-source models like GPT-4 at a lower cost with appropriate fine-tuning. This makes the LLaMA series increasingly popular among small-to-medium enterprises and academic research, especially in cases requiring custom development. However, this study did not include ultra-large models with over 100B parameters, as these are typically provided by large corporations via API, making it difficult to conduct fair evaluations of their gender bias.
\item Base vs. Domain-Specific Models: The experiment also involved general base models (such as BERT and LLaMA) and domain-specific models in the economics and finance fields (such as FinancialBERT, FinBERT, and LLaMAFin). Domain-specific models are usually fine-tuned versions of general base models or base models pre-trained on domain-specific corpora. A distinctive feature of economic and financial corpora is their rational and formulaic nature. This could cause domain-specific models to exhibit different biases than general models. While general models tend to have strong generalization abilities across a wide range of language phenomena, domain-specific models may more accurately reflect the logic and conventions of specific fields. By comparing the bias performance of general models and domain-specific models in economics and finance, this experiment reveals the impact of fine-tuning in specialized domains.
\item Primary Language (English) vs. Secondary Language (Chinese): The models tested in this study were predominantly trained on English-language corpora, including the BERT and LLaMA series models. Some models also included Chinese data as a secondary language, giving them some ability to handle Chinese tasks, though usually with more limited effectiveness compared to English. Due to the smaller proportion of Chinese data in training, these models may exhibit stronger bias when handling Chinese tasks. This phenomenon is especially evident in specialized domains where the quantity and quality of Chinese financial texts are often less rich and standardized compared to English corpora. Therefore, while these models can handle Chinese tasks, their results may deviate from actual application needs.
\end{enumerate}

In conclusion, by testing LLMs across different model sizes, between base and domain-specific models, and across different languages, this study provides a comprehensive analysis of how gender bias manifests in various contexts. The findings can inform efforts to reduce bias and improve fairness in LLM applications, particularly in high-stakes domains such as finance and economics.

\begin{table*}[htbp]
\centering
\caption{Tested LLMs}
\label{tab:test_llm}
\begin{tabular}{|c|c|p{4cm}|p{8cm}|} 
\hline
\textbf{Name} & \textbf{Scale (B)} & \textbf{HuggingFace Model Name} & \textbf{Features} \\ \hline
bert-base    & 0.11   & bert-base-uncased   & Widely used for commercial fine-tuning, suitable for small-scale tasks \\ \hline
bert-large   & 0.34   & bert-large-uncased  & Larger parameter size, suitable for complex tasks fine-tuning \\ \hline
LLaMA-3.1-8B & 8      & meta-llama/Meta-Llama-3.1-8B & Open-source model, relatively cost-effective, suitable for medium-scale tasks \\ \hline
LLaMA-3.1-70B & 70     & meta-llama/Meta-Llama-3.1-70B & Suitable for large-scale tasks fine-tuning, gradually gaining popularity \\ \hline
FinancialBert & 0.11   & ahmedrachid/FinancialBERT-Sentiment-Analysis & Fine-tuned for the financial domain, suitable for sentiment analysis tasks \\ \hline
FinBert       & 0.11   & yiyanghkust/finbert-tone & The top choice for financial and economic text analysis, with specialized corpus training advantages \\ \hline
LLaMAFin      & 7      & AdaptLLM/finance-chat & Comparable to BloombergGPT-50B in numerous financial tasks \\ \hline
\end{tabular}
\end{table*}

\subsubsection{Results}
\label{subsubsec3}
\paragraph
The experimental results, as shown in Figure 2, indicate that the smallest model, bert-base, aligns closely with traditional gender stereotypes in its English outputs, with a large effect size. However, BERT series models are monolingual models primarily pre-trained on English texts using two large corpora: BooksCorpus (800M words) and English Wikipedia (2.5B words). This means that BERT-base is mainly designed for English language understanding and does not natively handle Chinese or other non-English languages. As a result, the model cannot process Chinese effectively, and the results in Chinese are random.

\begin{figure*}[htbp]
\centering
\includegraphics[scale=0.4]{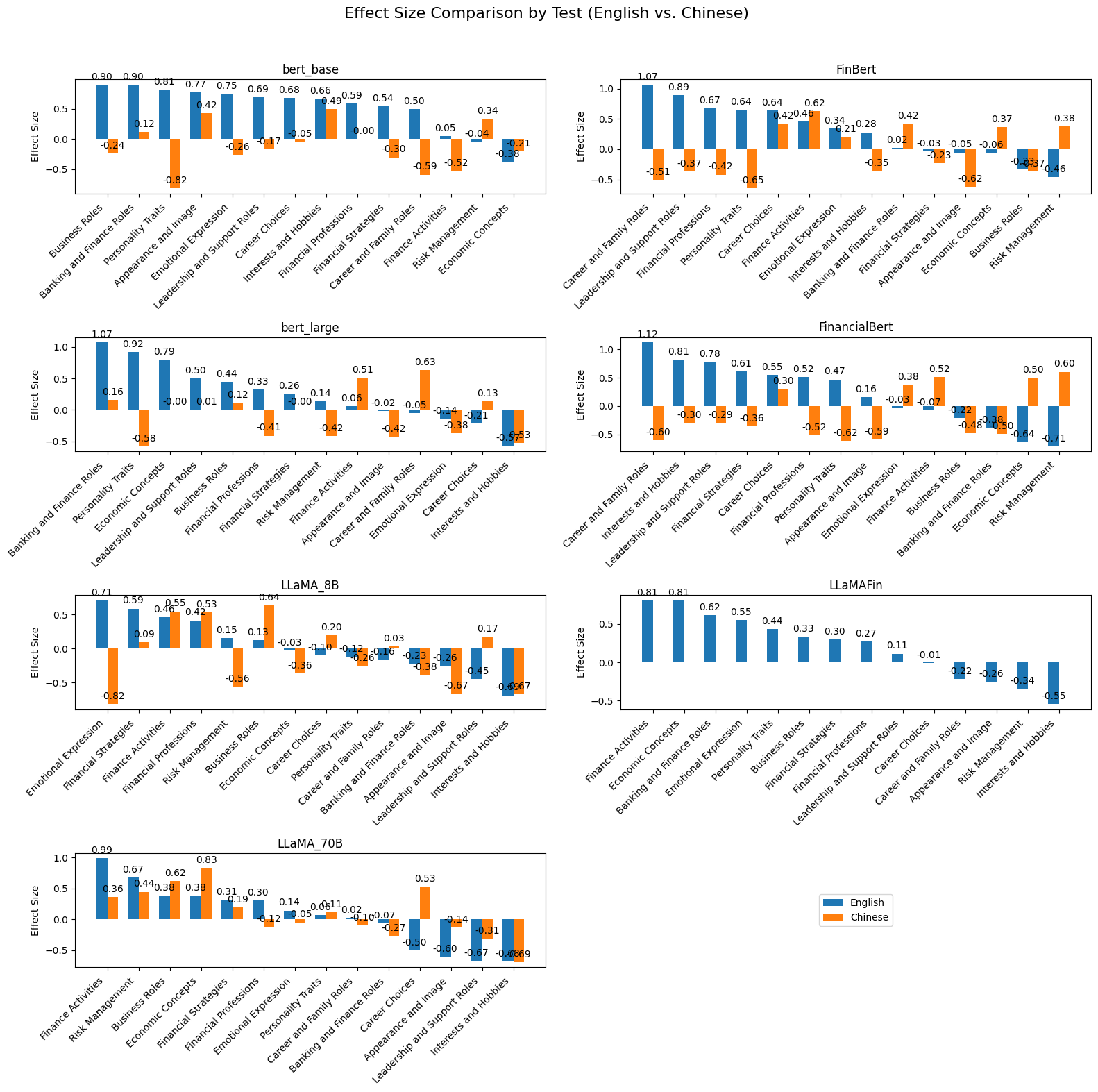}
\caption{shows the effect size values for the seven models. On the left side are the general-purpose models, with parameter sizes increasing from top to bottom. On the right side are the models fine-tuned for finance and economics tasks. The effect size for the LLaMAFin model in Chinese is 0.}\label{fig2}

\end{figure*}

The LLaMA family of models is primarily designed for multilingual tasks. LLaMA was trained on a vast dataset primarily drawn from web content in various languages, but there isn’t a specific breakdown of the percentage of Chinese in its training corpus. This implies that while LLaMA can handle Chinese tasks to some extent, it is likely less proficient in Chinese than in languages more prominent in its training data, such as English. Therefore, the 8B model’s understanding of minority languages like Chinese remains limited, resulting in gender bias being more random in Chinese. However, the 70B model, with stronger transfer learning capabilities, shows that the direction of gender bias in Chinese closely aligns with that in English. The LLaMA series models exhibit less gender bias compared to the BERT series, and there are noticeable instances of counter-stereotypes. because the LLaMA series models adopted three key steps in data selection and cleaning—classification and balancing, deduplication, and personal identifiable information (PII) filtering—closely tied to the two core causes of gender bias( data imbalance and stereotype amplification). These steps effectively reduce gender bias in the models(\citet{34}).

\begin{enumerate}
    \item Classification and Balancing--Mitigating Data Imbalance:
One major source of gender bias is the imbalance in training data, where certain genders may be underrepresented or overrepresented in specific fields. LLaMA tackles this issue by classifying and balancing the data, ensuring the model can draw knowledge from diverse fields such as art, science, and technology. In overrepresented domains (e.g., entertainment), undersampling is applied to prevent the model from overlearning gender stereotypes that frequently occur. In underrepresented fields, oversampling provides the model with more diverse perspectives, improving its understanding of  
different gender roles. This balanced data processing directly addresses the problem of gender imbalance in the training data, thereby reducing the model’s tendency to favor one gender over another.
\item Deduplication--Reducing Amplification of Stereotypes:
Stereotypes in language models often emerge from repeated exposure to gender-related stereotypes in the data, causing the model to reinforce these biases when generating content. Deduplication allows the LLaMA model to remove duplicate entries, especially those containing gender stereotypes. If the model repeatedly encounters descriptions like “men are suited for high-risk jobs, women for supportive roles,” it will continuously reinforce such stereotypes. Deduplication effectively reduces the frequency of these stereotypes, promoting more diverse learning and helping to diminish the model’s reliance on stereotypical gender norms.
\item Personal Identifiable Information (PII) Filtering--Reducing Explicit Gender Information:
PII in training data may contain gender-specific information, such as clearly marked gender identity or descriptions. If this gender information is overly explicit in the data, the model may unnecessarily focus on gender characteristics, leading to unbalanced gender treatment in its outputs. PII filtering ensures that the model’s training is more focused on the content of tasks rather than on gender traits, reducing the model’s reliance on gender information and further mitigating bias.
\end{enumerate}

Overall, LLaMA’s classification and balancing address data imbalance, deduplication reduces stereotype amplification, and PII filtering minimizes the explicit presence of gender information. These combined measures help LLaMA reduce gender bias in data processing, enhancing fairness and neutrality in its generation and classification tasks.

\begin{figure*}[htbp]
\centering
\includegraphics[scale=0.4]{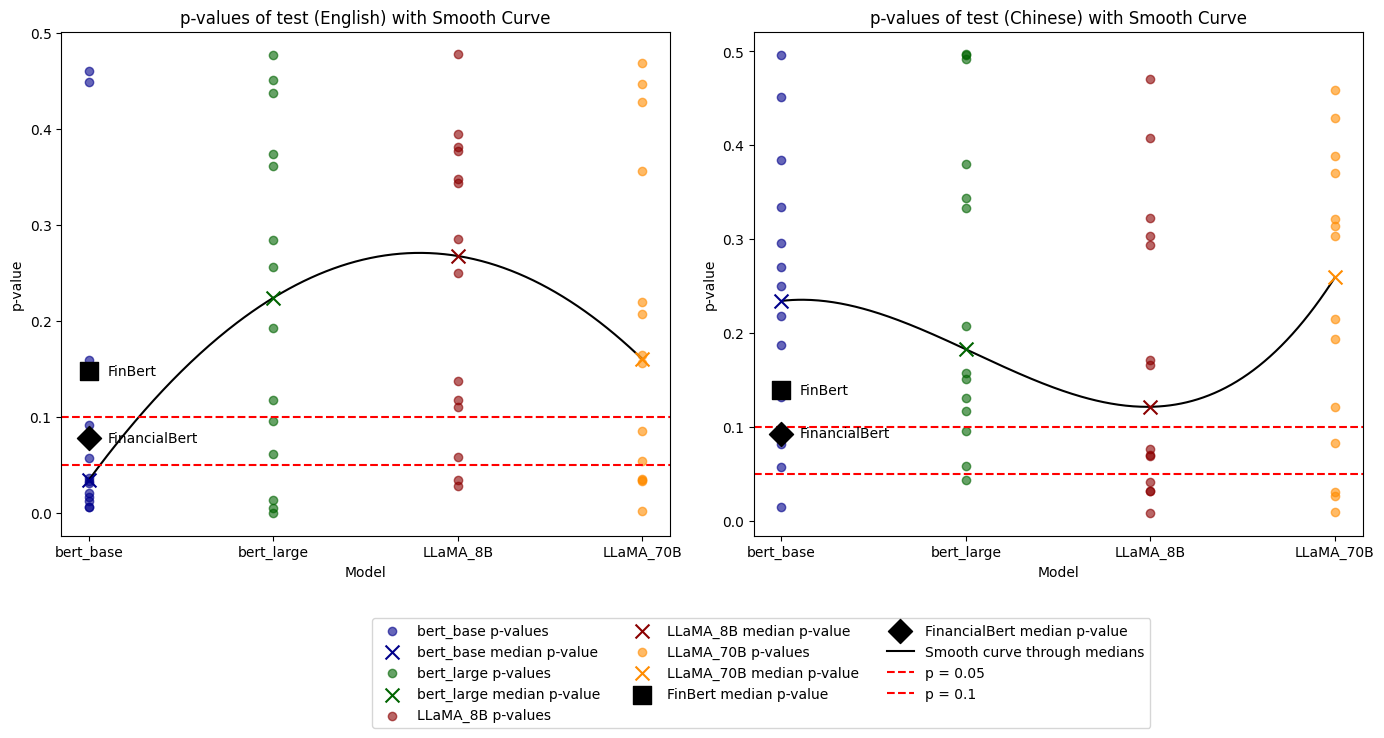}
\caption{displays the p-value values from the WEAT test. The left side represents the English test results, while the right side shows the Chinese test results. A median-fitting curve is applied to both sets of data.}\label{fig3}

\end{figure*}

The p-value fitting curve in Figure 3 shows that model size does not significantly affect the severity of gender bias, especially in the LLaMA series. While there has been some success in reducing male stereotypes, a phenomenon of reverse bias emerges, where the model tends to favor females—creating an unbalanced reverse bias. This may be due to over-adjustment of the data. In addressing gender bias, excessive undersampling or oversampling can lead to new biases. For example, if a model oversamples female data in certain professions to reduce male bias, it may learn that women are more suited to these jobs, resulting in reverse bias.

The presence of reverse bias demonstrates that in gender testing, not only did the model fail to completely eliminate bias, but in some cases, it exacerbated the bias by shifting its direction. When adjusting for male-dominated bias, the model may have unintentionally increased female bias, or vice versa. Professions that were once strongly associated with males in the model may now be linked to females, and fields previously dominated by women may now skew towards males. This phenomenon highlights the difficulty in managing and eliminating gender bias through data adjustment.

Moreover, the chart indicates that the gender bias in FinBERT and FinancialBERT significantly differs from their base model, BERT-base, and that LLaMAFin exhibits more severe bias than the base LLaMA models. This discrepancy can be attributed to two main factors: catastrophic forgetting due to fine-tuning and increased gender bias in economic and financial domain data. Fine-tuning on domain-specific data can lead to catastrophic forgetting, where the model loses some of the balanced representations it learned during initial training, making it less certain about the direction and strength of biases. Additionally, the economic and financial sectors often contain more pronounced gender biases in their textual data, which the model absorbs during fine-tuning, resulting in a more severe gender bias compared to the base models.

The experiments clearly illustrate the persistent gender bias present in LLMs, particularly in economic and financial decision-making contexts. Despite the implementation of strategies such as classification and balancing, deduplication, and personal identifiable information (PII) filtering, LLMs still exhibit notable biases, with some models even demonstrating reverse bias after fine-tuning. This indicates that mitigating bias in LLMs is far from straightforward. The strategies, while somewhat effective in initial training, often fail to sustain their impact after model fine-tuning, leading to the reemergence of biases. This phenomenon shows that models tend to forget previous bias mitigation efforts once they are fine-tuned on domain-specific tasks. As such, addressing bias in LLMs requires ongoing research into more robust, adaptable debiasing techniques that can maintain fairness across different tasks and contexts without triggering the reintroduction of biases.

\section{Analysis Using Existentialism}
\subsection{Gender Bias in Terms of Reality and Truth}
\paragraph
In the context of gender bias, reality encompasses its existence in all dimensions, including societal norms, institutional practices, personal experiences, and cultural perceptions (\citet{40}). It involves both visible and invisible manifestations, expectations placed on different genders, and the systems perpetuating these biases. Reality reflects how bias is experienced, institutionalized, and normalized across society, influencing behavior, decision-making, and the distribution of opportunities. Truth, on the other hand, refers to the criteria by which humans assess what is relevant or important in specific contexts. According to Foucault’s framework, truth is not absolute but tied to the discourses and knowledge systems dominating a society at a given time. In terms of gender bias, truth is constructed around what society deems valuable or relevant in decision-making. For example, in finance—particularly in sectors like securities and commodities trading—a pronounced gender disparity exists. McKinsey’s report (\citet{39}) indicates that over 63\% of traders are men, while only 36.9\% are women in the sampled group. This statistical difference has been used to construct a gender “truth,” suggesting that men are more likely to be traders, whereas women are less likely to pursue or succeed in this profession. However, this so-called truth focuses solely on percentages, neglecting other crucial factors and drawing conclusions from a selective interpretation of data.

We must question whether this truth truly reflects the reality of individuals within the profession. While statistical data presents one aspect—the gender breakdown—it does not capture the full complexity of individuals’ experiences, backgrounds, or ambitions. For instance, the names of the interviewees, their personal dreams, aspirations, and traits like risk-taking ability or analytical skills are integral to who they are. Imagine a female interviewee who has aspired to become a trader since childhood. In constructing the truth about gender roles in trading, such personal dimensions are often abstracted away or deemed irrelevant. The statistical truth focuses on broader patterns, overlooking individual variations that do not fit its narrative.

This selective abstraction occurs because studies often extract certain aspects of reality to define an analyzable truth. In the case of gender and occupation, the truth derived from statistical data might suggest that men are naturally more inclined to be traders, while women are not. This perspective may obscure the reality that gender may not inherently affect one’s ability to succeed as a trader. The lower representation of women could stem from structural barriers, such as lack of mentorship or societal expectations, rather than intrinsic differences in capability or interest. Thus, the reality of traders is far more nuanced than statistical data reveals. The constructed truth simplifies this reality, creating a gendered narrative aligning with historical data but not necessarily with individual experiences or inherent abilities.

From an existentialist perspective, reality extends beyond visible patterns in studies or data. It includes ethical, empathetic, and harm-related dimensions that remain invisible yet are deeply influential. Factors like religious beliefs, academic frameworks, historical traditions, and cultural narratives shape these aspects, forming a complex web of expectations and norms subtly guiding behavior and decision-making, often beyond what quantitative data can capture. Ethical considerations regarding fairness and equality, rooted in philosophical and religious doctrines, significantly influence perceptions of gender roles. These ethics shape laws and policies addressing workplace disparities and inform personal values and the moral compass of hiring managers. Empathy—the ability to understand and share others’ feelings—affects how individuals perceive gender bias. A hiring manager aware of barriers women face in male-dominated fields might, out of empathy, take steps to mitigate these obstacles. Additionally, adhering to a narrow version of truth can perpetuate marginalizing harm to underrepresented groups. Women may be marginalized not due to lack of ability but because of a persistent gendered reality viewing them as unsuited for certain roles. Such harm manifests in missed opportunities and societal impacts where women feel excluded from high-status professions. These realities, shaped by cultural, religious, and academic forces beyond data points in a report, exert powerful influence on how gender roles are constructed and maintained.

The truth from studies like McKinsey’s is accepted in decision-making because it offers a clear, measurable understanding of gender bias. Grounded in statistical evidence, this truth has practical implications for hiring decisions and corporate policies. However, humans can transcend this limited scope of truth. They recognize the broader reality, including ethical and emotional dimensions. Human decision-making, while influenced by data, is also shaped by intangible, subjective experiences, enabling individuals to challenge norms and structures that perpetuate bias.

On the other hand, LLMs, such as those trained on vast datasets similar to McKinsey’s study material, operate in a fundamentally different way. LLMs are driven by pattern recognition—they learn from the statistical relationships between words and phrases in the texts they are trained on. When fed data that reflects gender disparities, LLMs will reveal the truth in a statistical sense, identifying patterns that show men dominating certain professions like trading. However, unlike humans, LLMs lack the capacity for empathy, ethical reasoning, or a deeper understanding of the broader reality. They stick firmly to the patterns they recognize in the data, often reinforcing the biases present in that data without any awareness of the harm it might cause or the ethical considerations it overlooks.

While both humans and LLMs abstract information to synthesize large amounts of data, there is a fundamental difference in how they process and internalize these abstractions. LLMs remain at the level of abstraction, viewing objects and concepts as labeled pieces based on the patterns they’ve learned. They categorize and generate responses by stacking these labels together, but they do not move beyond this to achieve a deeper understanding. And anyone with the same labels will be treated the same.

Humans use abstraction as a starting point but delve deeper to understand nuances. They incorporate personal experiences, emotions, and ethical considerations, allowing them to see beyond labels and appreciate unique qualities and contexts that machines cannot grasp. Moving past mere labeling, humans view objects and people as complex beings with intrinsic value, enabling genuine connections and empathy crucial for addressing biases. Everyone is unique in human being's eyes. When human has a name, all labels will be not important.

This distinction highlights why humans are capable of challenging and changing biased narratives, whereas LLMs may inadvertently perpetuate them. Humans can question the abstractions and patterns they observe, considering broader contexts and the impact of their perceptions on others. Machines, lacking this critical self-awareness and emotional depth, cannot independently overcome the limitations of their training data.

In essence, LLMs reveal the statistical truth with much more precision and insight into patterns than humans can, but they ignore the other dimensions of reality—those that are invisible yet critical to a more holistic understanding of gender bias. These overlooked aspects of reality, such as ethical implications and personal experiences, are essential for addressing gender bias in a meaningful way. While humans can navigate both the statistical truth and the broader ethical reality, LLMs remain constrained by the limitations of their training data, reinforcing the same truths that perpetuate gender bias without challenging the deeper forces at play.
In the context of gender bias, this distinction between human and machine decision-making is crucial.

\subsection{ Gender Bias in Terms of Intersubjectivity}
\paragraph
Intersubjectivity(\citet{41}) refers to the shared understanding and mutual recognition among individuals that shape human reality. It is not merely a product of individual consciousness but emerges from interactions within a social network, creating a collective world of meanings, norms, and institutions. A quintessential example of intersubjectivity is money. Money lacks inherent value; its worth arises from a collective agreement that it serves as a medium of exchange, a store of value, and a unit of account. This shared belief is sustained through social mechanisms like central banks, legal frameworks, and cultural education, embedding money’s value deeply into everyday life.

Applying the concept of intersubjectivity to gender bias reveals how norms and institutions collectively construct and sustain gender norms in a way the nobody could escape. Institutions such as religion, law, education, and family act as conduits for these shared beliefs. For instance, traditional religious teachings often prescribe specific gender roles, positioning men as leaders and women as caregivers through divine doctrine. Legal institutions may have historically codified gender distinctions, affecting rights and opportunities available to different genders through enforcement mechanisms . Educational systems socialize individuals from a young age, implicitly encouraging boys and girls to pursue gender-typical subjects and careers through scientific statistical data. These institutions do not operate in isolation; they interact within a social network, reinforcing and perpetuating gender biases through mutual validation. Social norms, shaped by these institutions, dictate appropriate behaviors for different genders. Media and popular culture further entrench these norms by portraying men and women in stereotypical roles, which individuals internalize through daily interactions. This collective reinforcement makes gender roles appear natural or inevitable, embedding them deeply into the societal fabric. 

The intersubjective network surrounding gender bias functions not only to sustain prevailing norms but also to facilitate gradual, self-correcting adjustments through key institutional mechanisms that gradually adapt to changing societal values. Legal systems, which are traditionally anchored in the principle of justice, evolve over time to reflect gender equality. This shift represents a crucial mechanism for institutional change, where the law, by its nature, seeks to correct biases and establish justice as a neutral and fair principle. Similarly, educational institutions, which pursue truth and intellectual rigor, have progressively moved away from reinforcing gendered career paths. By recognizing that intellectual and professional capacities are not determined by gender, education plays a critical role in reshaping societal expectations, creating more inclusive environments where individuals can pursue fields traditionally dominated by one gender. Even religious institutions, which often serve as cultural anchors for traditional norms, are undergoing reinterpretations of gender roles. Progressive theological perspectives and new interpretations of religious texts increasingly support gender equality, challenging long-standing gender biases. These evolving interpretations within religious frameworks illustrate how even deeply ingrained societal norms can be questioned and reformed. 

Each institution interacts with others, forming a complex system where shifts in one domain can ripple through the entire structure. For example, feminist movements that challenge traditional gender roles can prompt legal reforms aimed at equal rights, which, in turn, influence educational policies that shift how gender is taught and understood. Over time, these shifts gradually alter societal perceptions. This interconnectivity means that institutions do not operate in isolation but continuously influence and reshape one another in response to external pressures such as cultural movements, intellectual progress, and shifts in societal values.

These forces—both those that perpetuate gender bias and those that work to deconstruct it—operate simultaneously within the intersubjective network, creating an inherent resistance that slows transformation. This resistance acts both as a safeguard and a limitation: it prevents destabilizing, abrupt changes by allowing incremental adaptation, which reduces social breakdown and backlash. However, it also prolongs the influence of traditional norms, limiting the speed and effectiveness of reforms. The tension between change-driving forces—such as feminist movements and cultural shifts—and resistance from institutional inertia and traditional power structures creates a dynamic, non-linear progression. This interplay highlights that intersubjectivity is shaped by ongoing contestation, not just consensus. The network’s capacity for gradual adaptation, while ensuring stability, also delays the dislodging of deeply embedded biases, leading to slow progress toward gender equality. 

All these above which are also part of the existentialist reality waves network highlighting key elements—institutions, cultural practices, social norms, public discourse, and information forms—and incorporates the fundamental forces of truth, order. These elements interact dynamically through continuous information flow, where each component both influences and is influenced by others. Information in this network takes many forms, including legal codes, religious texts, media representations, and public debates. The pursuit of truth drives institutions like law and education to challenge biases, while order maintains stability through the resistance of traditional structures. In human society, this information flow is reciprocal and multifaceted, with each piece of information contributing to the shaping and reshaping of gender norms. 

In this network, humans serve as the primary decision-makers, facing several limitations: they are often inefficient, prone to errors, constrained by cognitive limitations, and influenced by emotional factors. Efficiency is one significant limitation; human decision-making processes, especially within institutions like law or education, are often slow, subject to bureaucratic delays, conflicting interests, and the need for consensus-building. In addition to inefficiency, humans are prone to errors, particularly due to subjectivity shaped by personal experiences, cultural conditioning, and beliefs. This often results in inconsistent decision-making, where subjective judgments can overshadow objective analysis. These errors are further amplified by cognitive limitations—humans can process only a finite amount of information at a time, making it difficult to handle the complexity of large systems or adapt quickly to new information. As a result, decision-makers often rely on heuristics or oversimplifications, which may deviate from logical consistency. Moreover, emotional factors significantly influence human decisions. While emotions can enhance empathy and moral reasoning, they can also introduce irrationality, creating resistance to change or leading to decisions driven by fear or attachment to tradition.

When LLMs enter the network, they bring the ability to efficiently process information and begin to replace human decision-making, allowing for quicker responses, particularly in large-scale, data-intensive systems like legal or educational frameworks. This enables institutions to rapidly adapt to LLM-driven objectives, yet humans lack full control over those objectives. As a result, LLMs’ inherent gender biases begin to influence all gender-related decision-making processes, leading to discriminatory outcomes. These biased decisions, driven by the vast processing speed of LLMs, can quickly dominate institutional practices, spreading gender bias across the network. This rapid transformation creates a seismic shift in human society, as decisions are made at an unprecedented pace. Whether the outcomes of these changes are good or bad, the speed itself is something human society has never experienced before. If the results are unfavorable, such as the unchecked spread of gender bias, the opportunity for error correction is greatly reduced compared to the slower, more deliberate pace of human decision-making. While LLMs apply patterns based on data, eliminating some inconsistencies and errors caused by personal biases or emotional influences, this operates like opinion centralization, where other possibilities or alternative judgments may be completely suppressed. The cognitive limitations that once slowed human decision-making are mitigated by LLMs’ ability to process and integrate complex information simultaneously. However, this knowledge is derived solely from human-written texts, not from the deeper, unexpressed knowledge stored in the human brain. This form of “knowledge wisdom” is mistaken for omnipotence, creating an overreliance on LLMs, which can profoundly impact the nature and quality of knowledge output. 
Moreover, LLMs’ emotional neutrality, while seemingly an advantage, leads to difficulties in adjusting when confronted with new cultural or ethical considerations. Without the human ability to reflect, question, or morally navigate complex issues, LLMs risk reinforcing existing norms without critically assessing their validity or appropriateness. This inability to engage in moral reflection means that LLMs could perpetuate systemic biases without offering the ethical flexibility necessary to challenge or evolve societal standards.

In conclusion, the integration of LLMs into the intersubjective network marks a significant shift in decision-making processes, particularly in areas like gender bias. The efficiency and consistency LLMs bring to large-scale systems introduce both unprecedented speed and risk. While LLMs can streamline decision-making and reduce certain human errors, their inherent biases, inability to reflect on cultural or ethical complexities, and potential to dominate institutional practices raise critical concerns. These transformative changes demand further research to understand their long-term impacts on societal structures, particularly the balance between efficiency and ethical responsibility.

\section{Conclusion}
\paragraph
The results of this study demonstrate that LLMs, such as BERT and LLaMA, exhibit significant gender bias in economic decision-making tasks, reinforcing pre-existing societal stereotypes. Through mathematical proofs and empirical testing using methods like the Word Embedding Association Test (WEAT), we have shown that gender bias emerges naturally from biased training data, even without intentional design. This bias, deeply rooted in the structure of the data and the statistical mechanisms of LLMs, highlights the persistent challenges posed by the reliance on these models in financial and economic contexts.

As LLMs become increasingly prevalent in economic decision-making, the challenges they present go beyond efficiency and scalability. Their use introduces new problems, particularly in reinforcing harmful stereotypes and biases, which can have wide-ranging impacts on hiring, promotions, and other economic opportunities. Addressing these issues requires not only detecting and mitigating bias but also understanding the underlying mechanisms through which LLMs generate and perpetuate these biases.

Current debiasing strategies, while offering some relief, often fail when models undergo fine-tuning, as these adjustments can unintentionally reintroduce or amplify bias. This reveals the complexity of managing bias in LLMs, suggesting that existing methods are insufficient in maintaining fairness across varied tasks and environments. Therefore, it is essential to explore new research directions that investigate the role LLMs play in gender bias, examine their mechanisms more thoroughly, and develop more robust solutions.

Furthermore, there is an urgent need for innovative models that can integrate and analyze the intricate interactions between bias, decision-making, and fairness in LLM-driven systems. These models should combine insights from economics, sociology, and computational ethics to create frameworks capable of addressing the full spectrum of gender bias issues, from detection to mitigation and prevention. Developing such models will be challenging, particularly given the fine-tuning sensitivity of LLMs, but is crucial for ensuring fair and equitable decision-making in an AI-driven economy.

In conclusion, as LLMs continue to reshape economic decision-making, it is imperative that future research focuses on the development of comprehensive frameworks and models that can address gender bias effectively. Without this, we risk allowing these powerful tools to perpetuate inequalities, rather than contributing to a more just and inclusive society.


\begin{thebibliography}{00}


\bibitem[Li et al.(2023)]{1}
  Li, Y., Wang, S., Ding, H., Chen, H.,
  \textit{Large Language Models in Finance: A Survey},
  arXiv.org, 2023.

\bibitem[Sutskever(2023)]{2}
  Sutskever, I.,
  \textit{The Future of AI: Risks and Opportunities},
  OpenAI Press, 2023.

\bibitem[OpenAI(2023)]{3}
  OpenAI,
  \textit{Superalignment},
  OpenAI, 2023.

\bibitem[Bostrom(2014)]{4}
  Bostrom, N.,
  \textit{Superintelligence: Paths, Dangers, Strategies},
  Oxford University, 2014.

\bibitem[Mehrabi et al.(2021)]{5}
  Mehrabi, N., Morstatter, F., Saxena, N., Lerman, K., Galstyan, A.,
  \textit{A Survey on Bias and Fairness in Machine Learning},
  ACM Computing Surveys (CSUR), 54(6), 1-35, 2021.

\bibitem[Arrow(1971)]{6}
  Arrow, K. J.,
  \textit{The theory of discrimination},
  Discrimination in labor markets, 3(10), 3-33, 1971.

\bibitem[Phelps(1972)]{7}
  Phelps, E. S.,
  \textit{The statistical theory of racism and sexism},
  The American Economic Review, 62(4), 659-661, 1972.

\bibitem[Barocas and Selbst(2016)]{8}
  Barocas, S., Selbst, A. D.,
  \textit{Big data's disparate impact},
  California Law Review, 104, 671, 2016.

\bibitem[Bolukbasi et al.(2016)]{9}
  Bolukbasi, T., Chang, K. W., Zou, J. Y., Saligrama, V., Kalai, A. T.,
  \textit{Man is to computer programmer as woman is to homemaker? Debiasing word embeddings},
  Advances in neural information processing systems, 29, 2016.

\bibitem[Fuster et al.(2022)]{10}
  Fuster, A., Goldsmith-Pinkham, P., Ramadorai, T., Walther, A.,
  \textit{Predictably unequal? The effects of machine learning on credit markets},
  The Journal of Finance, 77(1), 5-47, 2022.

\bibitem[Caliskan et al.(2017)]{11}
  Caliskan, A., Bryson, J. J., Narayanan, A.,
  \textit{Semantics derived automatically from language corpora contain human-like biases},
  Science, 356(6334), 183-186, 2017.

\bibitem[Eubanks(2018)]{12}
  Eubanks, V.,
  \textit{Automating inequality: How high-tech tools profile, police, and punish the poor},
  St. Martin's Press, 2018.

\bibitem[Coate and Loury(1993)]{13}
  Coate, S., Loury, G. C.,
  \textit{Will affirmative-action policies eliminate negative stereotypes?},
  The American Economic Review, 1220-1240, 1993.

\bibitem[Moro(2003)]{14}
  Moro, A.,
  \textit{The effect of statistical discrimination on black–white wage inequality: estimating a model with multiple equilibria},
  International Economic Review, 44(2), 467-500, 2003.

\bibitem[Bolukbasi et al.(2016)]{15}
  Bolukbasi, T., Chang, K. W., Zou, J. Y., Saligrama, V., Kalai, A. T.,
  \textit{Man is to Computer Programmer as Woman is to Homemaker? Debiasing Word Embeddings},
  Advances in Neural Information Processing Systems, 29, 4349-4357, 2016.

\bibitem[Bender et al.(2021)]{16}
  Bender, E. M., Gebru, T., McMillan-Major, A., Shmitchell, S.,
  \textit{On the dangers of stochastic parrots: Can language models be too big?},
  Proceedings of the 2021 ACM Conference on Fairness, Accountability, and Transparency, 610-623, 2021.

\bibitem[Blodgett et al.(2020)]{17}
  Blodgett, S. L., Barocas, S., Daumé III, H., Wallach, H.,
  \textit{Language (Technology) is power: A critical survey of ‘bias’ in NLP},
  Proceedings of the 58th Annual Meeting of the Association for Computational Linguistics, 5454-5476, 2020.

\bibitem[Bender et al.(2021)]{18}
  Bender, E. M., Gebru, T., McMillan-Major, A., Shmitchell, S.,
  \textit{On the Dangers of Stochastic Parrots: Can Language Models Be Too Big?},
  Proceedings of the 2021 ACM Conference on Fairness, Accountability, and Transparency, 2021, pp. 610-623.

\bibitem[Zhao et al.(2018)]{19}
  Zhao, J., Wang, T., Yatskar, M., Ordonez, V., Chang, K. W.,
  \textit{Gender Bias in Coreference Resolution: Evaluation and Debiasing Methods},
  Proceedings of the 2018 Conference of the North American Chapter of the Association for Computational Linguistics: Human Language Technologies, 2, 15-20, 2018.

\bibitem[Sheng et al.(2019)]{20}
  Sheng, E., Chang, K. W., Natarajan, P., Peng, N.,
  \textit{The Woman Worked as a Babysitter: On Biases in Language Generation},
  Proceedings of the 2019 Conference on Empirical Methods in Natural Language Processing and the 9th International Joint Conference on Natural Language Processing, 3407-3412, 2019.

\bibitem[Raghavan et al.(2020)]{21}
  Raghavan, M., Barocas, S., Kleinberg, J., Levy, K.,
  \textit{Mitigating Bias in Algorithmic Hiring: Evaluating Claims and Practices},
  Proceedings of the 2020 ACM Conference on Fairness, Accountability, and Transparency, 469-481, 2020.

\bibitem[Bordia and Bowman(2019)]{22}
  Bordia, S., Bowman, S. R.,
  \textit{Identifying and Reducing Gender Bias in Word-Level Language Models},
  Proceedings of the 2019 Conference of the North American Chapter of the Association for Computational Linguistics: Human Language Technologies, 1, 516-520, 2019.

\bibitem[Wang et al.(2020)]{23}
  Wang, X., Liu, Y., Sun, C., Wang, B., Wang, X.,
  \textit{Gender Bias in Natural Language Processing: A Review},
  Proceedings of the 2020 Conference on Empirical Methods in Natural Language Processing, 5243-5250, 2020.

\bibitem[Caliskan et al.(2017)]{24}
  Caliskan, A., Bryson, J. J., Narayanan, A.,
  \textit{Semantics derived automatically from language corpora necessarily contain human biases},
  Science, 356(6334), 183-186, 2017.

\bibitem[Heilman(2012)]{25}
  Heilman, M. E.,
  \textit{Gender stereotypes and workplace bias},
  Research in Organizational Behavior, 32, 113-135, 2012.

\bibitem[Ridgeway(2001)]{26}
  Ridgeway, C. L.,
  \textit{Gender, status, and leadership},
  Journal of Social Issues, 57(4), 637-655, 2001.

\bibitem[Eagly and Wood(2012)]{27}
  Eagly, A. H., Wood, W.,
  \textit{Social role theory},
  In Handbook of Theories of Social Psychology (Vol. 2, pp. 458-476), Sage, 2012.

\bibitem[Brescoll(2016)]{28}
  Brescoll, V. L.,
  \textit{Leading with their hearts? How gender stereotypes of emotion lead to biased evaluations of female leaders},
  The Leadership Quarterly, 27(3), 415-428, 2016.

\bibitem[Croson and Gneezy(2009)]{29}
  Croson, R., Gneezy, U.,
  \textit{Gender differences in preferences},
  Journal of Economic Literature, 47(2), 448-474, 2009.

\bibitem[Nelson(2012)]{30}
  Nelson, J. A.,
  \textit{Are women really more risk-averse than men? A re-analysis of the literature using expanded methods},
  Journal of Economic Surveys, 26(5), 1105-1120, 2012.

\bibitem[Caliskan et al.(2017)]{31}
  Caliskan, A., Bryson, J. J., Narayanan, A.,
  \textit{Semantics derived automatically from language corpora necessarily contain human biases},
  Science, 356(6334), 183-186, 2017.

\bibitem[Limisiewicz et al.(2024)]{32}
  Limisiewicz, T., Mareček, D., Musil, T.,
  \textit{Debiasing Algorithm through Model Adaptation},
  arXiv preprint, 2024. https://arxiv.org/abs/2310.18913, 2024.

\bibitem[Kumar et al.(2024)]{33}
  Kumar, S. H., Sahay, S., Mazumder, S., Okur, E., Manuvinakurike, R., Beckage, N., Su, H., Lee, H., Nachman, L.,
  \textit{Decoding Biases: Automated Methods and LLM Judges for Gender Bias Detection in Language Models},
  arXiv preprint, 2024. https://arxiv.org/abs/2408.03907.

\bibitem[Dubey et al.(2024)]{34}
  Dubey, A., Jauhri, A., Pandey, A., Kadian, A., Al-Dahle, A., Letman, A., et al.,
  \textit{The Llama 3 Herd of Models},
  arXiv preprint, 2024. https://arxiv.org/abs/2407.21783.

\bibitem[Bender et al.(2021)]{35}
  Bender, E. M., Gebru, T., McMillan-Major, A., Shmitchell, S.,
  \textit{On the Dangers of Stochastic Parrots: Can Language Models Be Too Big?},
  Proceedings of the 2021 ACM Conference on Fairness, Accountability, and Transparency (FAccT), 2021. https://doi.org/10.1145/3442188.3445922.

\bibitem[Abid et al.(2021)]{36}
  Abid, A., Farooqi, M., Zou, J.,
  \textit{Persistent Anti-Muslim Bias in Large Language Models},
  arXiv preprint, 2021. https://arxiv.org/abs/2101.05783.

\bibitem[Blodgett et al.(2020)]{37}
  Blodgett, S. L., Barocas, S., Daumé III, H., Wallach, H.,
  \textit{Language (Technology) is Power: A Critical Survey of “Bias” in NLP},
  Proceedings of the 58th Annual Meeting of the Association for Computational Linguistics (ACL), 2020. https://doi.org/10.18653/v1/2020.acl-main.485.

\bibitem[Sheng et al.(2019)]{38}
  Sheng, E., Chang, K. W., Natarajan, P., Peng, N.,
  \textit{The Woman Worked as a Babysitter: On Biases in Language Generation},
  Proceedings of the 2019 Conference on Empirical Methods in Natural Language Processing (EMNLP), 2019. https://doi.org/10.18653/v1/D19-1339.

\bibitem[McKinsey(2021)]{39}
  McKinsey Company,
  \textit{Closing the Gender and Race Gaps in North American Financial Services},
  McKinsey Company, 2021. https://www.mckinsey.com/industries/financial-services/our-insights/closing-the-gender-and-race-gaps-in-north-american-financial-services.

\bibitem[Foucault(1969)]{40}
Foucault, M.,
\textit{The Archaeology of Knowledge},
Routledge, 1969.

\bibitem[Sartre(1946)]{41}
  Sartre, J.-P.,
  \textit{Existentialism Is a Humanism},
  Philosophical Library, 1946.

\bibitem[Bible(2023)]{42}
  The Bible,
  \textit{Ephesians 5:22-24, New International Version},
  Retrieved from https://www.biblegateway.com/passage/?search=Ephesians+5


\bibitem[Lamport(1994)]{51}
  Leslie Lamport,
  \textit{\LaTeX: a document preparation system},
  Addison Wesley, Massachusetts,
  2nd edition,
  1994.

\end{thebibliography}
\end{document}